\documentclass[aps,prb,superscriptaddress,showpacs,showkeys,preprint,12pt]{revtex4-1}
\usepackage{amssymb,amsmath,amsbsy,bm,graphicx,hyperref,xcolor,csquotes,braket,wasysym,verbatim}

\begin{document}
\title{Confined electron states in two-dimensional HgTe in magnetic field: Quantum-dot versus  quantum-ring behavior}

\author{Du\v{s}an B. Topalovi\'c}\email{dusan.topalovic@vin.bg.ac.rs}
\affiliation{School of Electrical Engineering, University of Belgrade, P. O. Box
35-54, 11120 Belgrade, Serbia} \affiliation{Vin\v{c}a Institute of Nuclear Sciences,
University of Belgrade, P. O. Box 522, 11001 Belgrade, Serbia}
\author{Vladimir V. Arsoski}\email{vladimir.arsoski@etf.bg.ac.rs}
\affiliation{School of Electrical Engineering, University of Belgrade, P. O. Box
35-54, 11120 Belgrade, Serbia}
\author{Milan \v{Z}. Tadi\'c}\email{milan.tadic@etf.bg.ac.rs}
\affiliation{School of Electrical Engineering, University of Belgrade, P. O. Box
35-54, 11120 Belgrade, Serbia}
\author{Fran\c{c}ois M. Peeters}\email{francois.peeters@uantwerpen.be}
\affiliation{School of Physics and Astronomy and Yunnan Key Laboratory for
Quantum Information, Yunnan University, Kunming 650091, China}
\affiliation{Department of Physics, University of Antwerp, Groenenborgerlaan
171, B-2020 Antwerp, Belgium}

\begin{abstract}
We investigate the electron states and optical absorption in square- and hexagonal-shaped two-dimensional (2D) HgTe quantum dots and quantum rings in the presence of a perpendicular magnetic field. The electronic structure is modeled by means of the $sp^3d^5s^*$ tight-binding method within the nearest-neighbor approximation. Both bulklike and edge states appear in the energy spectrum. The bulklike states in quantum rings exhibit Aharonov-Bohm oscillations in magnetic field, whereas no such oscillations are found in quantum dots, which is ascribed to the different topology of the two systems. When magnetic field varies, all the edge states in square quantum dots appear as quasibands composed of almost fully flat levels, whereas some edge states in quantum rings are found to oscillate with magnetic field. However, the edge states in hexagonal quantum dots are localized like in rings. The absorption spectra of all the structures consist of numerous absorption lines, which substantially overlap even for small line broadening. The absorption lines in the infrared are found to originate from transitions between edge states. It is shown that the magnetic field can be used to efficiently tune the optical absorption of HgTe 2D quantum dot and quantum ring systems.
\end{abstract}

 \maketitle

\section{Introduction}\label{I}

In the past few decades, low-dimensional nanometer-sized structures have attracted a lot of attention because of their unique properties and potential applications in electronics and photonics.\cite{Delerue04, Harrison05, Woggon97} Depending on the employed fabrication method, nanostructures have different dimensionality and are generally classified as quantum wells, quantum wires, and quantum dots. Among them, quantum dots (QDs) are systems in which the electrons and holes are confined in all three spatial directions. They feature a discrete energy spectrum and have tunable optical properties. Systematic study of QDs commenced with the works of Brus in the early 1980s.\cite{Brus83,Brus84} In the ensuing years, a few methods were developed for the fabrication of QDs, including colloidal chemical synthesis,\cite{Guzelian96,Dabbousi97} hybrid chemical-electrochemical synthesis,\cite{Penner00} lithography,\cite{Bertino07} and molecular beam epitaxy.\cite{Nakata00,Petroff94,Alchalabi03}

HgTe QDs have been fabricated and used for various optoelectronic devices which operate in the infrared range of the optical spectrum.\cite{Connor05, Keuleyan11} They are usually synthesized by the colloidal growth technique,\cite{Rogach99} and are therefore called colloidal quantum dots (CQDs). Due to quantum confinement the exciton energy increases when the CQD size decreases and the energy bands become inverted. Hence, a negative band gap could be achieved which is important for infrared applications. Optical properties of HgTe CQDs could be tuned by temperature,\cite{Lhuillier12} and CQDs can be designed such that sharp absorption lines from near- ($\lambda\approx$ 1.3 $\mu$m) to mid-infrared ($\lambda\approx$ 5 $\mu$m) are present.\cite{Keuleyan11b}

Also, HgTe has recently gained popularity due to its unique topological properties. It is known that the valence band of HgTe is $s$-like and has $\Gamma_6$ symmetry, while the conduction band is $p$-like and has $\Gamma_8$ symmetry. When bands are inverted a quantum phase transition from an ordinary insulator to a topological insulator (TI) occurs. It has been reported that monolayer HgTe achieves a topological nontrivial phase under in-plane tensile strain.\cite{Li15} Furthermore, HgTe quantum wells exhibit both a band gap and robust topological gapless edge states which are protected by time-reversal symmetry.\cite{Bernevig06, Konig07}

Phonon calculations confirmed that two-dimensional (2D) HgTe honeycomb layers are dynamically stable and may be used for the design of novel 2D heterojunction devices.\cite{Zheng15} These honeycomb monolayers are not perfectly planar and they have a low-buckled configuration. Motivated by these studies, we investigate electronic states and optical absorption of 2D HgTe square quantum dots (SQDs), hexagonal quantum dots (HQDs), square quantum rings (SQRs) and hexagonal quantum rings (HQRs) in a perpendicular magnetic field. Besides QDs we investigate quantum rings (QRs) because of enhanced confinement due to the presence of internal edges. This causes the emergence of new interesting effects in a magnetic field, which are primarily related to the occurrence of Aharonov-Bohm (AB) oscillations.\cite{Chakraborty18, Yeyati95, Alfonso05, Grbic08} The AB oscillations in the electronic spectra of thin HgTe quantum rings embedded in an $n$-type HgTe/HgCdTe quantum well were experimentally investigated more than a decade ago.\cite{Konig06,Konig07b} However, those rings were mesoscopic, having an average radius of the order of 1 $\mu$m \cite{Konig06}, for which small Aharonov-Bohm oscillations were measured.\cite{Konig06,Konig07b} Nanoscopic HgTe rings, whose average radius is of the order of a few nanometers, have not been produced by now. However, no obstacle would hinder fabrication of such nanorings by means of state-of-the-art nanotechnology. HgTe CQDs were fabricated as small as 10 nm in radius, and we believe that similarly such small HgTe quantum rings can be achieved in the near future. The electron states in nanostructures could be substantially modified by varying their geometrical parameters.\cite{Liang15, Cantele01, Holtkemper18, Zhu97}

In this paper, we investigate different possibilities of controlling the properties of HgTe 2D QDs and QRs through the control of shape and type of edges. We analyze theoretically whether the electron states in HgTe S(H)QDs and S(H)QRs are similarly arranged. Section II presents our model, which is based on the tight-binding (TB) method, and we show how an external magnetic field is included in the model. The results of the calculations for both energy levels and the absorption spectra are presented and discussed in Sec. III. Finally, our findings are summarized and a concise conclusion is given in Sec. IV.

\section{Theoretical model}

\begin{figure} \centering
\includegraphics[width=8.6cm]{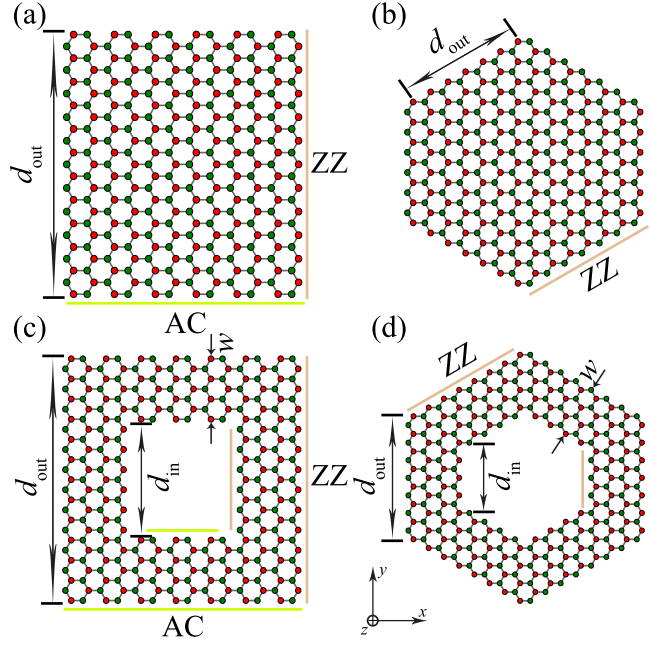}
\caption{In-plane schematic view of numerically investigated (a) SQD,  (b) HQD, (c) SQR, and (d) HQR: $d_{\rm out}$ is the side length of a polygonal QD and the outer side of QRs, $d_{\rm in}$ is the length of the inner side of QRs, and $w$ is the ring width. AC and ZZ highlight armchair and zigzag types of boundaries, respectively. Green (red) circles represent Te (Hg) atoms.}
\label{fig1}
\end{figure}

Figure \ref{fig1} shows a schematic view of the studied 2D nanostructures positioned in the $x-y$ plane. We denote by $d_{\rm out}$ the side length of polygonal QDs (outer side of QRs); $d_{\rm in}$ is the length of the inner side of QRs, and $w$ is the ring width. In square structures we adopted mixed armchair (AC) and zigzag (ZZ) boundary conditions, while hexagonal structures have only ZZ edges. All nanostructures have equal surfaces $S = 25.30\:\rm{nm}^2$. Table I summarizes geometrical parameters ($d_{\rm out}, d_{\rm in}, w$) and number of edge atoms ($N_{\Sigma,\rm ea}$).

\begin{table}
\caption{Review of shape, type, geometrical parameters ($d_{\rm out}, d_{\rm in}, w$), and number of edge atoms ($N_{\Sigma,\rm ea}$) for studied nanostructures: $d_{\rm out}$ denote either edge length of QDs or outer edge of QRs, $d_{\rm in}$ is length of the inner edge of QR and $w$ is ring width. AC and ZZ denote armchair and zig-zag edge, respectively.}
\begin{tabular}{l*{5}{c}}
	\hline
	\hline
	Shape/Type & Boundary & $d_{\rm{out}}(d_{\rm{in}})$ (nm) & $w$ (nm) & $N_{\Sigma,\rm{ea}}$ \\
	\hline
	$\Square$/Dot & AC + ZZ & 5.03(-) & - & 46 \\
	$\varhexagon$/Dot & ZZ & 3.12(-) & - & 40 \\
	$\Square$/Ring & AC + ZZ & 5.63(2.53) & 1.55 & 76 \\
	$\varhexagon$/Ring & ZZ & 3.52(1.63) & 1.89 & 68 \\
	\hline
	\hline
\end{tabular}\label{tab:I}
\end{table}

In order to model the electron states in HgTe QDs and QRs we employed the TB Hamiltonian \cite{Slater54, Shen12}	
\begin{equation}
	  H = \sum_{i, \sigma = \uparrow, \downarrow} \epsilon_{i} c^{\dagger}_{i, \sigma} c_{i, \sigma} - \sum_{\langle i, j \rangle, \sigma = \uparrow, \downarrow} t_{i, j} c^{\dagger}_{i, \sigma} c_{j, \sigma}.\label{eq:Hamiltonian}
\end{equation}
Here, $i$ runs over the atomic sites in the crystal lattice, $j$ runs over the nearest neighbors to atom $i$, and $\sigma$ denotes the electron spin up ($\uparrow$) and spin down ($\downarrow$), respectively. $\epsilon_i$ is the matrix of the on-site energies of atom $i$, and the intrinsic spin-orbit interaction (SOI) is included through nondiagonal terms.\cite{Chadi77} Two such matrices are computed, for a Te anion and for a Hg cation separately. The off-site matrix $t_{i,j}$ describes the electron hopping from site $j$ to site $i$.\cite{Slater54} $c^{\dagger}_{i, \sigma}$ and $ c_{i, \sigma}$ are the creation and annihilation operators of the electron with spin $\sigma$ at site $i$. The basis of $sp^3d^5s^*$ orbitals was used in the TB approximation and hopping was limited to the first nearest neighbor. Atoms with less than two bonds have been removed. The values of on-site energies and hopping terms are taken from Ref.~\onlinecite{Allan12}. Moreover, in the absence of external magnetic field the eigenenergies are double degenerate, which is a consequence of time-reversal symmetry described by the Kramers theorem.\cite{Klein52}

The influence of the magnetic field is included through the Peierls substitution which modifies the off-site matrix by introducing a field-dependent phase shift \cite{Graf95,Dumitrica98,Pertsova15,Chen19}
\begin{equation}
	t_{i, j} \rightarrow t_{i, j} e^{i \frac{2\pi}{\Phi_{0}} \int\limits_{\mathbf{r_{i}}}^\mathbf{{r_{j}}} \mathbf{A} d \mathbf{r}},\label{eq:Peierls}		
\end{equation}
where $\Phi_0 = h/e$ is the flux quantum, $h$ is the Planck constant, $e$ is the elementary charge, and $\mathbf{A}$ is the magnetic vector potential. Up to first approximation, when magnetic field effects on an intra-atomic scale are neglected, this modification is gauge invariant.\cite{Graf95,Boykin01b} We used the Landau gauge $\mathbf{A} = (0, Bx, 0)$ where magnetic field is directed along the $z$ axis. For the hoppings from site $i$ to sites $j$ only the limits of the integral in (\ref{eq:Peierls}) are exchanged, hence $t_{i,j}=t_{j,i}^\dagger$. When spin is included, eigenenergies for spin-up (spin-down) states are decreased (increased) proportional to the magnetic field value.\cite{Graf95,Shanavas15} Since this has no physical significance in explaining considered effects, interaction of the external magnetic field with the electron spin magnetic dipole moment is not included in the model. \cite{Pertsova15,Chen19} The eigenenergies and eigenvectors are found numerically by means of the PYBINDING package,\cite{MoldovanPB} which diagonalizes Eq.\:(\ref{eq:Hamiltonian}).

The electronic density of states (DOS) of a QD system is represented by a sum of delta functions centered at the eigenenergies of the system. QDs are usually fabricated in ensembles, where the QD size cannot be fully controlled. This size variation leads to inhomogeneous broadening of spectral lines of optical absorption or emission. In order to describe this broadening, the expression for DOS is modified, 
\begin{equation}
	{\rm{DOS}}(E) = \sum_{n} \Gamma(E_{n} - E),
\end{equation}
following Refs.~\onlinecite{Yates07,Li17}. Here, $E_n$ is the $n$th state energy and $\Gamma$ is the Gaussian function,
\begin{equation}
	\Gamma(\Delta E) = \frac{1}{\sqrt{2\pi}c} e^{-\frac{(\Delta E)^{2}}{2c^{2}}}.
\end{equation}
Such a modified expression for DOS substantially eases the numerical calculation of optical absorption. Moreover, the broadening factor $c$ in such a calculation is usually taken to be constant for all the bands.\cite{Yates07}
  
The dipole matrix element ${\mathbf{P}}_{i,j}=<j|{\mathbf{r}}|i>$ between the states $i$ and $j$ is determined by including only intraorbital terms, which was demonstrated to be a reasonable approximation in recent calculations of the absorption spectra of quantum dots made of phosphorene,\cite{Li17} silicene, and graphene.\cite{Abdelsalam16} The absorption $A$ depends on the matrix element, and takes into account energy conservation \cite{Abdelsalam16}
\begin{equation}
	A(\hbar\omega) = \sum_{i, j} (E_{j} - E_{i}) | \bm{\varepsilon} \mathbf{P}_{i, j}|^2 \Gamma(E_{i} - E_{j} +\hbar\omega).	
\end{equation}
Here $\hbar\omega$ denotes the photon energy, $E_i$ and $E_j$ are the eigenenergies of the initial and final states, respectively, and ${\bm{\varepsilon}}$ is the polarization vector of light.

\section{Results and discussion}
\subsection{Influence of geometry, type of nanostructure and imposed boundary conditions}

\begin{figure} \centering
\includegraphics[width=16cm]{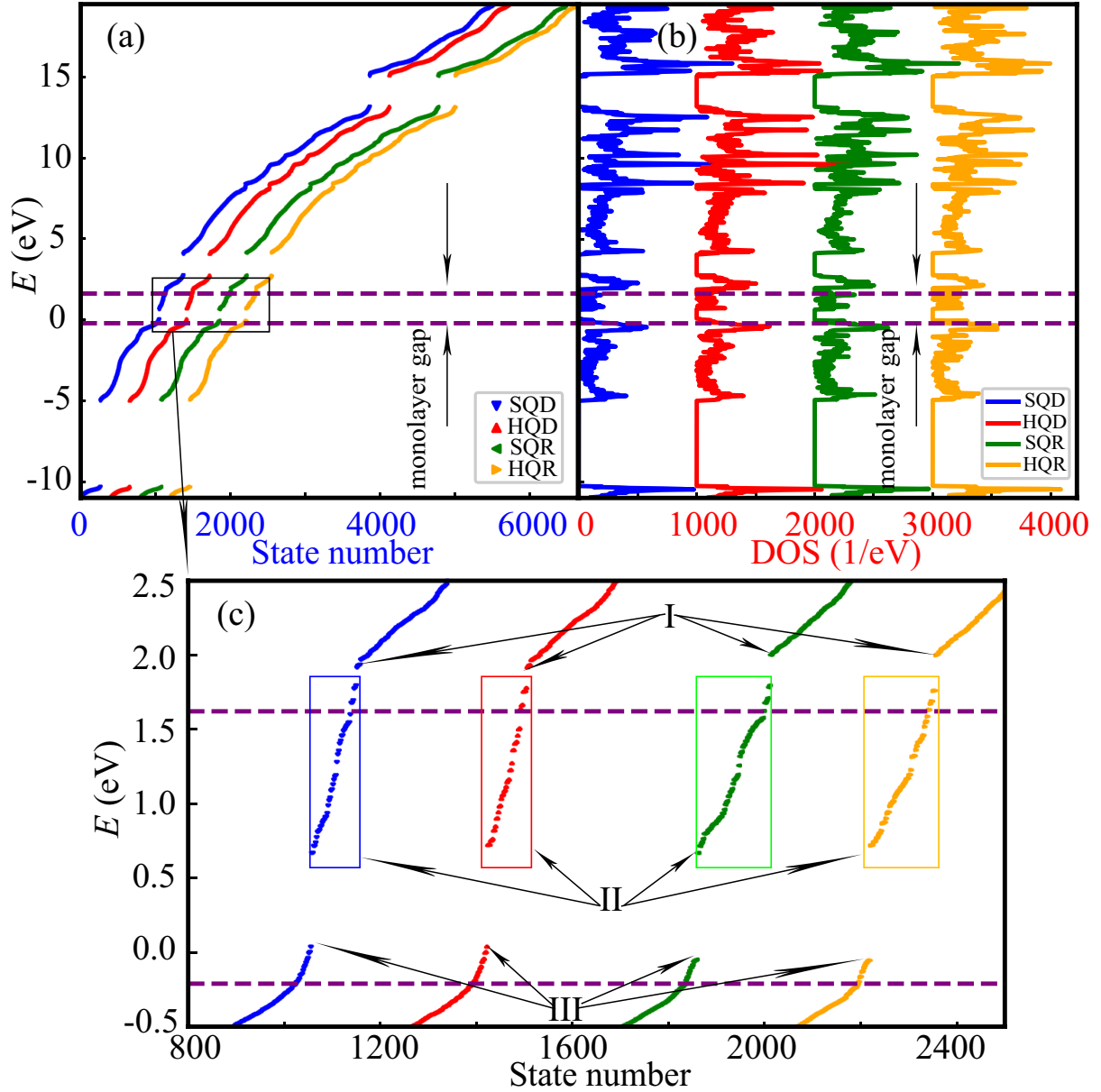}
\caption{(a) The distributions of the energy levels in the S(H)QDs and the S(H)QRs with the state number $n$. The different results are displaced by $\Delta n=400$. The energy window corresponding to the energy gap in 2D HgTe is denoted by horizontal purple dashed lines. (b) Variations of the DOS in the S(H)QDs and the S(H)QRs with energy. The different results are displaced by 1000 eV$^{-1}$. (c) A detailed view of the energy levels displayed in (a). The electron density for a few of the selected states denoted by I, II, and III, are shown in Fig.\:\ref{fig3}.}
\label{fig2}
\end{figure}

The edge lengths of the analyzed polygonal QDs and QRs given in Table~\ref{tab:I} correspond to recently fabricated HgTe QD samples, whose dimensions were found to vary in the range from 4 to 15 nm.\cite{Keuleyan11}  The eigenenergies in the analyzed QDs and the QRs at $B = 0$ are displayed in Fig.\:\ref{fig2}(a). The energy dependence of the DOS for the four types of structures is computed with a broadening of $c = 10$ meV\cite{Li17} and is shown in Fig.\:\ref{fig2}(b).

Figure \:\ref{fig2}(a) shows that the energy levels in all four systems vary similarly with the state number. We examine in more detail a part of the spectrum inside the fundamental band gap of bulk 2D HgTe. It occurs between the $\Gamma_4$ conduction band and the $\Gamma_{5,6}$ valence bands.\cite{Li15} This band gap is 1.82 eV wide and is delineated by horizontal purple dashed lines in Fig.\:\ref{fig2}(a). In HgTe QDs and QRs no continuous bands exist, but the discrete states are organized like in bands. We found that five such {\it quasibands} take place from -11 eV to 19.5 eV. Here, the energy origin is the valence-band maximum in 2D HgTe. Energy gaps between the quasibands have the property of vanishing  DOS, and because of quantum confinement they are wider than the fundamental band gap (see Table~\ref{tab:II}). Also, the DOS variation with energy exhibits sharp peaks close to the quasiband extrema, as shown in Fig.\:\ref{fig2}(b). The DOS peaks in the four structures do not differ much in either magnitude or location. But they differ slightly in the range of the fundamental band gap of the 2D system, as Fig.\:\ref{fig2}(c) shows. This energy window is populated by edge states in all four structures, which is similar to that found for silicene, germanene, and phosphorene quantum dots and quantum rings.\cite{Jakovljevic17, Li17} Such states are labeled by II in Fig.\:\ref{fig2}(c). Even though edge states exist in all structures, they exhibit slight differences, which will be further illustrated and is of importance for the optical properties of the analyzed systems.

\begin{table}
\caption{The values of the energy gap (in units of eV) in the S(H)QD and the S(H)QR.}	
	\begin{tabular}{l*{4}{c}}
		\hline
		\hline
		$E_{\mathrm{g, SQD}}$ & $E_{\mathrm{g, HQD}}$ & $E_{\mathrm{g, SQR}}$ & $E_{\mathrm{g, HQR}}$ \\
		\hline
		1.875 & 1.876 & 2.046 & 2.048 \\
		\hline
		\hline
	\end{tabular}\label{tab:II}
\end{table}

\begin{figure} \centering
\includegraphics[width=16cm]{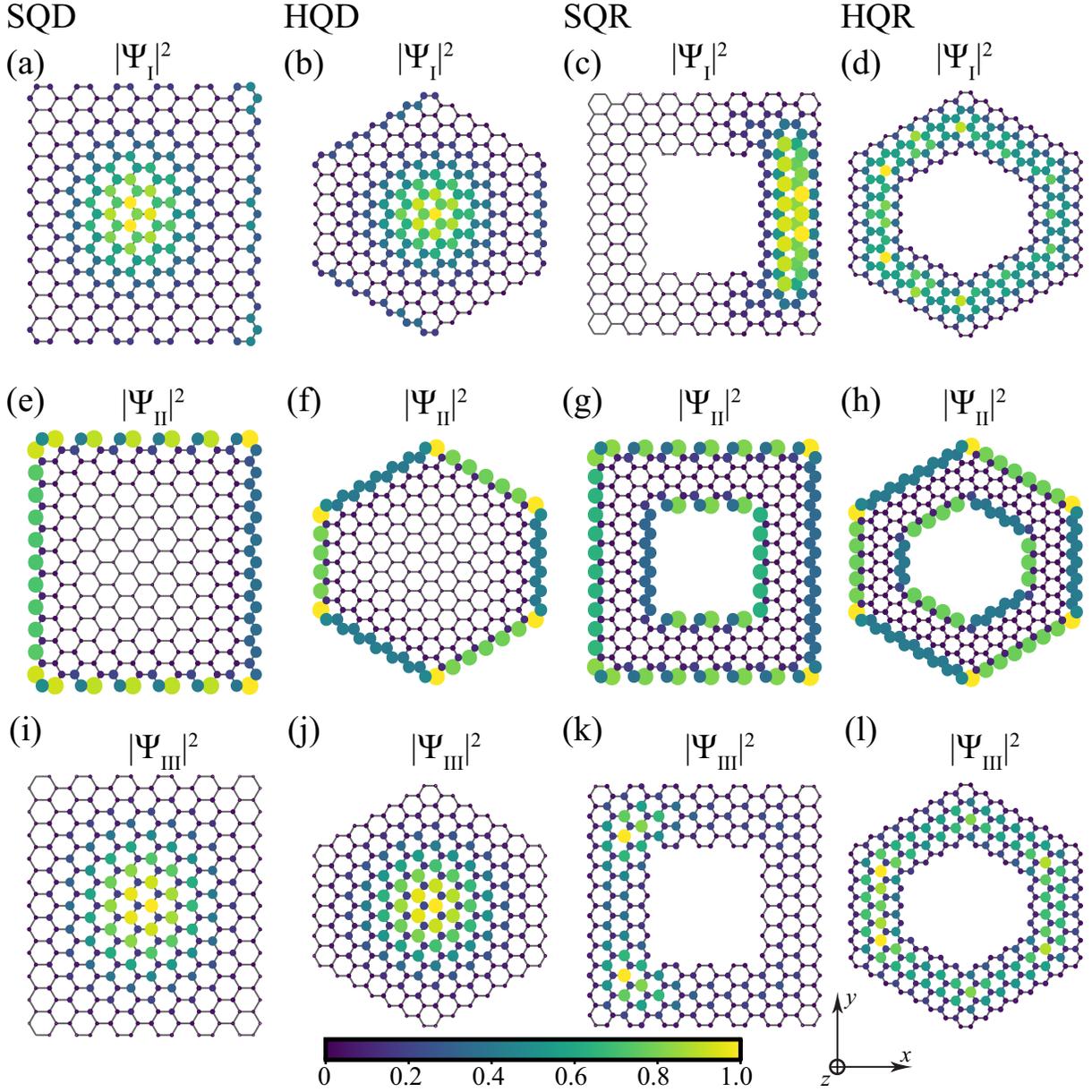}
\caption{The probability densities of the conduction-band minimum, the edge states, and valence-band maximum states in the S(H)QDs and the S(H)QRs labeled by I, II, and III in Fig.\:\ref{fig2}(c) at zero magnetic field. Top: The probability density of the lowest-energy conduction-band state $|\Psi_{\rm I}|^2$ in (a) the SQD, (b) the HQD, (c) the SQR, and (d) the HQR. Middle: The sum of the probability densities $|\Psi_{\rm II}|^2$  of the edge states in (e) the SQD, (f) the HQD, (g) the SQR, and (h) the HQR. Bottom: The probability density of the highest-energy valence-band state $|\Psi_{\rm III}|^2$ in (i) the SQD, (j) the HQD, (k) the SQR, and (l) the HQR.}
\label{fig3}
\end{figure}

In order to illustrate the different types of electron localization, we show in Fig.\:\ref{fig3} a few states in the quasibands close to the band gap of the 2D HgTe, which are denoted by I and III in Fig.\:\ref{fig2}(c). All the states have energies inside the band gap of the 2D HgTe. The upper panel in Fig.\:\ref{fig3} displays the probability density $|\Psi_{\rm I}|^2$ of the lowest-energy conduction-band state in the SQD, the HQD, the SQR, and the HQR. The sum of the probability densities  $|\Psi_{\rm II}|^2$ of the edge states is displayed in the middle panel (Figs.\:\ref{fig3}(e)-\:\ref{fig3}(h)), whereas the lower panel (Figs.\:\ref{fig3}(i)-\:\ref{fig3}(l)) shows the probability density $|\Psi_{\rm III}|^2$ of the highest-energy valence-band state. It appears that the electron states in the conduction and valence bands are situated throughout the dot (ring) and are therefore called {\it bulklike} states. On the other hand, the states shown in Figs.\:\ref{fig3}(e)-\:\ref{fig3}(h) are {\it{edge states}}, whose wave functions are confined to the edges of the dot or ring. We previously mentioned that edge states have energies inside the energy gap of 2D HgTe; thus, they are in-gap states. It is easy to show that the number of such states is equal to the sum of Hg and Te atoms situated at the edges. Hence, the number of edge states in rings is larger than the number of edge states in similar-sized quantum dots. Edge states are mainly localized at the sites of the Te atoms in all the cases (see middle panel of Fig.\:\ref{fig3}). This is a consequence of the larger Pauling electronegativity of Te than of Hg.\cite{Zheng15} Moreover, we notice in Fig.\:\ref{fig3} that the edge states are localized at both the ZZ and AC boundaries.

\subsection{Influence of a perpendicular magnetic field}

\begin{figure} \centering
\includegraphics[width=16cm]{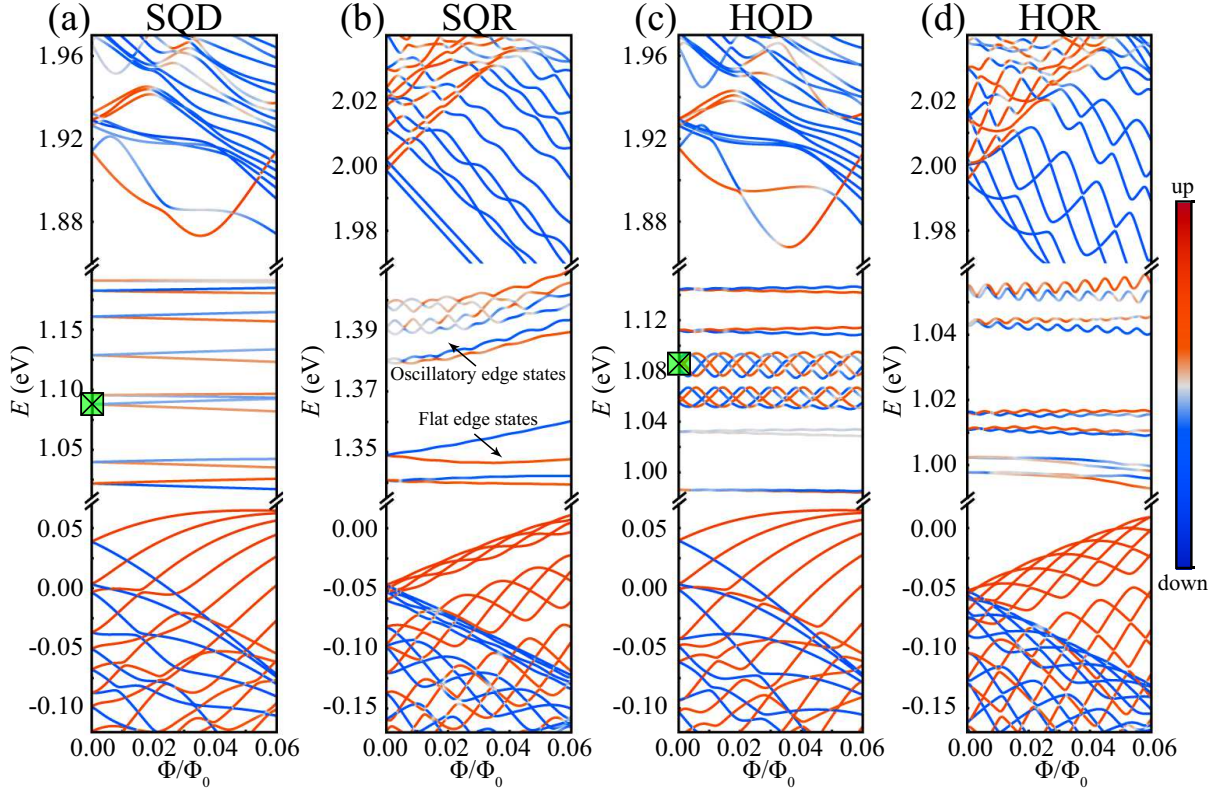}
\caption{The energy levels as function of perpendicular magnetic field in (a) the SQD, (b) the SQR, (c) the HQD, and (d) the HQR. The upper panel shows the electron states at the bottom of the conduction band, the middle panel displays the edge states, and the lower panel shows the electron states at the top of the valence band. The magnitudes of the spin-up or the spin-down components of the wave functions are indicated by coloring the energy levels with different shades of red and blue, respectively.}
\label{fig4}
\end{figure}

Figure \ref{fig4} demonstrates how the energy spectra are affected by a perpendicular magnetic field. Similar to previous figures, Figs.\:\ref{fig4}(a) and \:\ref{fig4}(b) depict the states in the SQD and the SQR, respectively, whereas Figs.\:\ref{fig4}(c) and \:\ref{fig4}(d) show the states in the HQD and the HQR, respectively. The energy levels are here shown as they vary with magnetic field flux $\Phi$ throughout the surface area of the 2D HgTe primitive cell divided by the magnetic flux quantum $\Phi_0$. The primitive cell is the same for all the considered morphologies, and such defined $\Phi_0$ provides a consistent approach suitable for comparisons of different structures. This approach is often used (see, for example, Ref.~\onlinecite{Li17}), even though $\Phi_0$ corresponds to a rather large value of the magnetic field ($\Phi/\Phi_0=B/22 936$ T).

Time-reversal symmetry is in this case broken; hence the spin degeneracy is lifted and the wave function of each state is a superposition of the spin-up and spin-down wave functions. Nonetheless, if the spin-up part of the total wave function is larger than the spin-down part, we refer to such a mixed state as a spin-up state. On the other hand, if the spin-down part is larger the state is called the spin-down state. When magnetic flux varies, the energy of the states of the opposite spins exhibits anticrossings,\cite{Jakovljevic17} with the dominant spin component being exchanged between the states that anticross. In a large quantum dot, however, the potential is much less effective in confining the electron than the magnetic field. In this case the energy levels converge to the Landau levels of the 2D system,\cite{Grujic11, Li17} and it indeed appears for magnetic fields beyond the range in Fig.~\ref{fig4}.  Furthermore, as Fig.\:\ref{fig4}(a) shows, bulklike and edge states vary differently with magnetic field. As a matter of fact, the electron wave functions decay in an area much smaller than the magnetic length $l_B=\sqrt{\hbar/eB}$, and thus the magnetic field is not effective in shifting the energy levels. As an example, we take a fairly large $B=10$ T for which $l_B=5.73$ nm, whereas an edge state is localized over a distance of the order of the HgTe lattice constant. Consequently the energies of the edge states do not vary much with magnetic field, and the diagram of energy dependence on magnetic field consists of almost flat lines. Also, Zeeman spin splitting is small in the range of magnetic field shown in Fig.\:\ref{fig4}.

\begin{figure} \centering
\includegraphics[width=16cm]{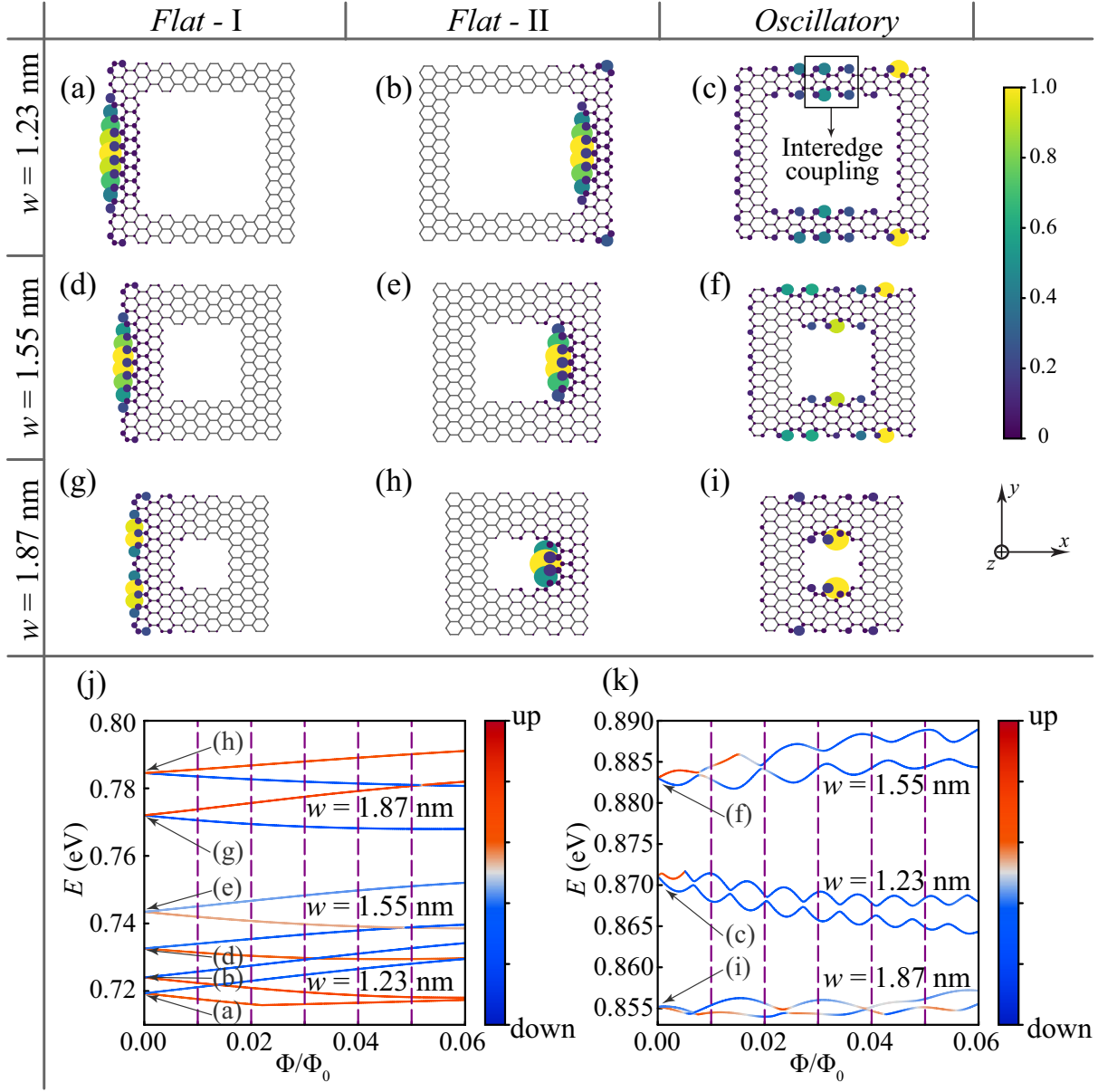}
\caption{The probability density at zero magnetic field for two flat edge states (denoted by {\it Flat}\:-\:I and {\it Flat}\:-\:II) and one oscillatory edge state in (a, c) 1.23-, (d, f) 1.55-, and (g, i) 1.87-nm-wide SQR. (j) The energy levels for the {\it Flat}\:-\:I and {\it Flat}\:-\:II edge state as a function of perpendicular magnetic field in 1.23-, 1.55-, and 1.87-nm-wide rings. The spin-up(down) states are shown by red(blue) line. (k) The energy levels for the {\it oscillatory} edge state as function of perpendicular magnetic field in 1.23-, 1.55-, and 1.87-nm-wide rings. The spin-up (spin-down) states are shown by the red (blue) line.}
\label{fig5}
\end{figure}

The magnetic field dependence of the states in the SQR has a few important differences from that in the SQD states, which could be inferred from a comparison between Figs.\:\ref{fig4}(a) and \:\ref{fig4}(b). First, the bulklike states in the SQR exhibit AB oscillations, which is not a feature of the SQD. The period of these oscillations approximately corresponds to one flux quantum threading a one-dimensional circular ring that encloses the area $A_{eff}=(d_{in}^2+d_{out}^2)/2$. Second, some of the edge states in the SQR are considerably affected by magnetic field, exhibiting similar oscillations to bulklike states. And the period of these oscillations strongly depends on the width and size of the ring.

We analyze rings that occupy an approximately equal area $S$ by Hg and Te atoms, but have different width. In Figs.\:\ref{fig5}(a)-\:\ref{fig5}(i) we display the probability density of the representative spin-split doublets in 1.23- (Figs.\:\ref{fig5}(a)-\:\ref{fig5}(c)), 1.55- (Figs.\:\ref{fig5}(d)-\:\ref{fig5}(f)), and 1.87-nm-wide (Figs.\:\ref{fig5}(g)-\:\ref{fig5}(i)) SQRs. One may notice that the diagrams are not symmetric with respect to rotations by either $\pi/2$ or $\pi$ rad around the axis perpendicular to the ring through the ring center. For all three values of the ring width the wave function of the state shown in the left-hand panel (see Figs.\:\ref{fig5}(a), \:\ref{fig5}(d), and \:\ref{fig5}(g)) is localized at the left part of the zigzag edge. On the other hand, the state whose eigenenergy is closest to the one shown in the left-hand panel of Fig.\:\ref{fig5} has a wavefunction localized mainly along the right inner edge (see Figs.\:\ref{fig5}(b), \:\ref{fig5}(e), and \:\ref{fig5}(h)). Hence, the electron wave function does not circulate around the edge regardless of the ring width. It in turn leads to almost flat energy states when magnetic field varies, which is demonstrated in Fig.\:\ref{fig5}(j). The states shown in this figure could be called {\it flat edge states}. On the other hand, the diagrams of the probability density shown in the right-hand panel (Figs.\:\ref{fig5}(c), \:\ref{fig5}(f), and \:\ref{fig5}(i)) are slightly delocalized with the wave functions spreading confined to more than a single edge. In this case the eigenenergies oscillate with magnetic field, as Fig.\:\ref{fig5}(k) shows. Because the electron in this state is localized at both the outer and the inner edges, the sides of the ring edge which face each other across a ring lobe might be considered to represent a coupled system. One may note in Fig.\:\ref{fig5} that the coupling is stronger if the ring is narrower, and that the edge state becomes more delocalized. The period of the oscillations increases with the ring width and is approximately $0.007\Phi_0$, $0.011\Phi_0$, and $0.013\Phi_0$ for 1.23-, 1.55-, and 1.87-nm-wide rings, respectively. Since all the rings contain an approximately equal number of atoms, one might conclude that the narrower the ring the larger the area it surrounds, and magnetic flux which threads the ring increases. Thus, the period of oscillations of the edge states decreases. The period of the oscillations of the bulklike states decreases in a similar fashion to the edge states as the ring width varies. Furthermore, the oscillations have a more regular shape in narrow quantum rings, which is clearly demonstrated in Fig.\:\ref{fig5}(k). Moreover, the area enclosed by the wave function of the edge states is somewhat larger than for bulklike states. Therefore, the period of the AB-like oscillations of the edge states is somewhat smaller than for the bulklike states.

\begin{figure} \centering
\includegraphics[width=8.6cm]{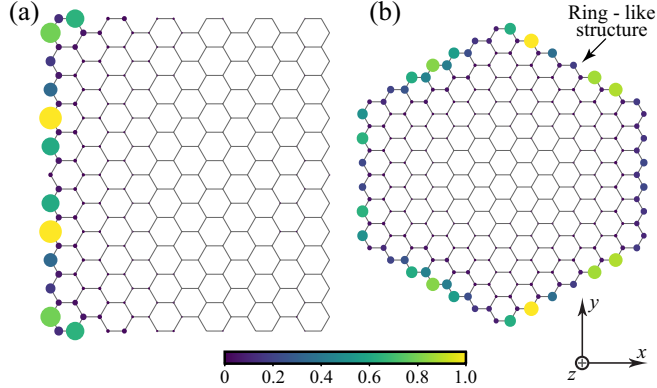}
\caption{The probability density of (a) a flat edge state in the SQD and (b) an oscillating edge state in the HQD. The energies of the displayed states are labeled by green squares symbols in Figs.\:\ref{fig4}(a) and \:\ref{fig4}(c).}
\label{fig6}
\end{figure}

The HQD and the HQR eigenenergies' dependence on magnetic field is shown in Figs.\:\ref{fig4}(c) and \:\ref{fig4}(d), respectively. By comparing Fig.\:\ref{fig4}(a) with Fig.\:\ref{fig4}(c) we see that bulklike states in the SQD and the HQD exhibit qualitatively similar dependence on $B$. This is a consequence of single-connectedness and almost equal areas of both systems. Also, the states in the HQD for even larger magnetic field are found to tend to Landau levels when magnetic field increases. Nonetheless, the edge states in the HQD show a few peculiar properties. Some of them exhibit oscillations with $B$ which resemble oscillations of the edge states in the SQR. This could be explained by comparing the states at zero magnetic field labeled by green squares in Figs.\:\ref{fig4}(a) and \:\ref{fig4}(c), and whose wave functions are shown in Figs.\:\ref{fig6}(a) and \:\ref{fig6}(b), respectively. From Fig.\:\ref{fig6}(a) we see that the edge state in the SQD is localized along only one ZZ side of the square, whereas the wave function in the HQD shown in Fig.\:\ref{fig6}(b) is distributed around the whole hexagonal edge. Such a wave function that is fully extended over the dot edge is responsible for the appearance of AB-like oscillations in the HQD, whereas oscillating edge states are not present in the SQD.

The case of the HQR depicted in Fig.\:\ref{fig4}(d) is similar to the case of the SQR shown in Fig.\:\ref{fig4}(b), thus demonstrating that the behavior of both bulklike and edge states in a magnetic field is mainly related to the topology. In the HQR we found that the wave functions of all the states are localized close to the edges and that the bulklike states exhibit AB oscillations. Also, the edge states in the HQR have wave functions which are either localized at certain spots at the edge or delocalized over the whole edge. The edge states whose wave functions are delocalized over the whole edge are evidently oscillatory with magnetic field. The oscillations of both the bulklike and oscillatory edge states in the HQR are more regular than in the SQR (compare Figs.\:\ref{fig4}(b) and \:\ref{fig4}(d)). The HQR shape indeed does not deviate much from a circular ring, where oscillations occur with a fixed period.

\subsection{Optical absorption in the presence of a perpendicular magnetic field}

\begin{figure} \centering
\includegraphics[width=8.6cm]{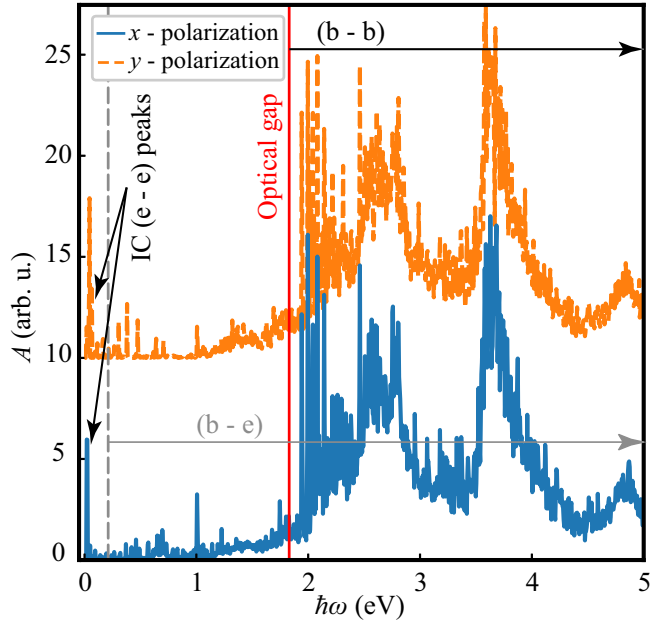}
\caption{The optical absorption spectrum of the SQD in the absence of external fields for $x$ and $y$ polarizations. The results for $y$ polarization are shifted up by 10 arb. units. The red vertical line denotes the fundamental gap while the gray vertical dashed line denotes the minimum energy of the b-e transitions. The range of the b-b and b-e transitions and the dominant e-e peaks for the two polarizations are indicated. The Fermi level is $E_F = 0.9$ eV.}
\label{fig7}
\end{figure}

We found that the electron states in 2D HgTe QDs and QRs show certain peculiarities which are related to material properties, structure shape, and topology. The electron states in turn will affect the optical absorption. Since all four structures have bulklike and edge states, the optical absorption could be classified to arise from (i) transitions between the bulklike states, denoted by b-b; (ii) transitions between the bulklike and edge states, abbreviated by b-e; and (iii) e-e transitions which take place between the edge states. We assume that the Fermi level approximately is situated in the middle of the energy gap between the valence and conduction bulklike states, and equals $E_F = 0.9$ eV for the SQD. Thus, it intersects the quasiband of edge states. Figure \:\ref{fig7} demonstrates that the highest absorption peak arises from the transitions between states across the Fermi level. Here we assume that the absorption lines are broadened by 2 meV, as in Ref.~\onlinecite{Abdelsalam16}, and $T = 0$ K is taken for temperature.

We calculate the optical absorption in the absence of external field for light polarized along both the $x$ and $y$ axes. For both polarization directions absorption is computed in the range of photon energies from 0 to 5 eV. For zero magnetic field the optical absorption spectra for all four structures have similar shapes. The qualitative similarities are a consequence of (1) the high in-plane symmetry of the structures, (2) the position of the Fermi level in the middle of the valence and conduction quasibands, (3) nearly equal numbers of atoms and edge states in all four structures, and (4) localization of the edge states on both the ZZ and AC edges. Hence, all four structures exhibit similar e-e optical absorption for both  polarizations. Let us compare the optical absorption in the SQD in the absence of a magnetic field shown in Fig.\:\ref{fig7} with the absorption spectra of phosphorene dots computed in Ref.\:\onlinecite{Li17}. There are some differences which are mainly a consequence of the different band gaps of the two materials. The band gap of phosphorene is large and the edge states are far from both the conduction and valence quasibands. Therefore, e-e, b-e and b-b transitions are well separated. In the analyzed HgTe quantum dots and rings, however, the edge states are rather close to the valence and the conduction bands. Hence, the absorption lines originating from the different types of transitions overlap with each other. Nevertheless, the optical absorption gap can still be resolved. It roughly separates the ranges where the b-b peaks and the peaks arising from the b-e and e-e transitions take place (see Fig.\:\ref{fig7}). Due to the similar localization of the conduction and valence bulklike states, which leads to large overlap between the wave functions, the b-b absorption peaks have higher amplitude as compared to the b-e and e-e ones. And, quite high peaks due to e-e transitions take place at about 25 meV for $x$ polarization and about 50 meV for $y$ polarization of incoming light.

The displayed absorption spectra show noticeable differences from the experimental results previously published for the three-dimensional (3D) HgTe CQDs, even though the respective dimensions of the two were similar.\cite{Keuleyan11b} The CQDs have stacked many layers of 2D HgTe;\cite{Keuleyan11} thus, their band gap is lower than the band gap of the quantum dots formed out from a monolayer of HgTe. Because the wave function in the 3D case extends normal to the HgTe multilayers, the low-energy absorption edge in the CQD is lower than the lowest energy of the b-b transition in the analyzed quantum dots. As a matter of fact, the transitions which take place between the edge states and the other states in the 2D system extend the absorption spectrum to even 50 $\mu$m. This value is substantially larger than the experimental absorption edge of the CQDs, which amounts to 5 $\mu$m for the dot average diameter of 10.5 nm.\cite{Keuleyan11}

\begin{figure} \centering
\includegraphics[width=8.6cm]{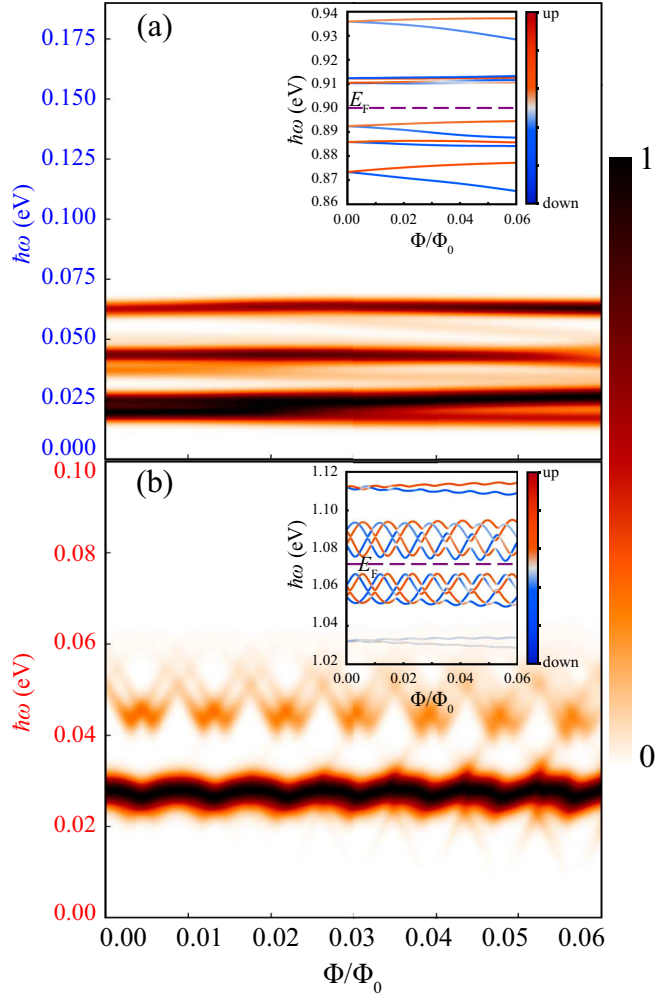}
\caption{A contour plot of the absorption amplitude that arises from transitions between (a) the flat edge states in the SQD, and (b) the oscillatory edge states in the HQD as function of magnetic field and absorption energy. The absorption is here computed as the average for $x$ and $y$ polarizations, and is scaled such that the maximum equals unity. The insets show the parts of the energy spectra around the Fermi level that significantly contribute to the absorption spectrum.}
\label{fig8}
\end{figure}

Finally, we focus on the magnetic field dependence of the e-e transitions in the SQD and HQD. The absorption spectra of these two structures differ considerably, as shown in Figs.\:\ref{fig8}(a) and (b). Here, the absorption is computed as an average between the values for the two polarizations and it is scaled such that the maximum equals unity. Evidently, the transition energies vary with magnetic field similar to the energies of the edge states. In the SQD the transitions are between flat edge states; thus, the absorption peaks in Fig.\:\ref{fig8}(a) almost do not shift with magnetic field. On the other hand, the transitions between the oscillatory edge states in the HQD are shown in Fig.\:\ref{fig8}(b). The transition energies corresponding to the absorption peaks oscillate with magnetic field. Also, the magnitude of the absorption is found to be highly sensitive to the magnetic field, and the e-e absorption might be considerably reduced when magnetic field is varied.

\section{Summary and conclusions}

The eigenenergies and the wave functions in square and hexagonal quantum dots and quantum rings made of 2D HgTe were computed by using the $sp^3d^5s^*$ basis set in the tight-binding model. The discrete states in the analyzed systems are densely distributed in ranges called quasibands, and are classified as either bulklike or edge states. The bulklike states with opposite spins exhibit anticrossings in magnetic field, whereas the edge states are mainly localized at the sites of the Te atoms. The energy of the edge states in {\it square} quantum dots almost does not depend on the magnetic field. On the other hand, some edge states in quantum rings are delocalized throughout the ring, and their energy levels oscillate with magnetic flux. This effect is similar to the Aharonov-Bohm oscillations of bulklike states in quantum rings, but the period of the oscillations of the edge states was found to be shorter. This could be explained by the smaller effective area bounded by the wave function of the bulklike state than the area which is encircled by the wave function of an edge state. Such oscillations were also found in {\it hexagonal} quantum dots. The transitions between the conduction- and valence-band bulklike states are found to dominate the absorption spectra of the analyzed structures. But in infrared there exists a relatively high absorption peak due to transitions between the edge states. Moreover, we found that the oscillatory edge states exhibit oscillatory variation of the position and amplitude of the absorption peaks with magnetic field.

\begin{acknowledgments}This work was supported by Projects No. III 41028, No. III 42008, and No. III 45003 funded by the Serbian Ministry of Education, Science and Technological Development, and the Flemish Science Foundation (FWO-Vl).
\end{acknowledgments}


\begin{thebibliography}{00}

\bibitem{Delerue04}C. Delerue and M. Lannoo, {\it Nanostructures: Theory and Modelling}, (Springer, Berlin, 2004), pp. 141-191 and 235-270.
\bibitem{Harrison05}P. Harrison, {\it Quantum wells, wires and dots: Theoretical and Computational Physics of Semiconductor Nanostructures}, (Wiley, Chichester, 2005), pp. 443-457.
\bibitem{Woggon97}U. Woggon, {\it Optical Properties of Semiconductor Quantum Dots}, (Springer, Berlin, 1997), pp. 223-230.
\bibitem{Brus83}L. E. Brus, J. Chem. Phys. {\bf 79}, 5566 (1983).
\bibitem{Brus84}L. E. Brus, J. Chem. Phys. {\bf 80}, 4403 (1984).
\bibitem{Guzelian96}A. A. Guzelian, U. Banin, A. V. Kadavanich, X. Peng, and A. P. Alivisatos, Appl. Phys. Lett. {\bf 69}, 1432 (1996).
\bibitem{Dabbousi97}B. O. Dabbousi, J. Rodriguez - Viejo, F. V. Mikulec, J. R. Heine, H. Mattoussi, R. Ober, K. F. Jensen, and M. G. Bawendi, J. Phys. Chem. B {\bf 101}, 9463 (1997).
\bibitem{Penner00}R. M. Penner, Acc. Chem. Res {\bf 33}, 78 (2000).
\bibitem{Bertino07}M. F. Bertino, R. R. Gadipalli, L. A. Martin, L. E. Rich, A. Yamilov, B. R. Heckman, N. Leventis, S. Guha, J. Katsoudas, R. Divan, and D. C. Mancini, Nanotechnology {\bf 18}, 315603 (2007).
\bibitem{Nakata00}Y. Nakata, K. Mukai, M. Sugawara, K. Ohtsubo, H. Ishikawa, and N. Yokoyama, J. Cry. Growth {\bf 208}, 93 (2000).
\bibitem{Petroff94}P. M. Petroff and S. P. DenBaars, Superlattices Microstruct. {\bf 15}, 15 (1994).
\bibitem{Alchalabi03}K. Alchalabi, D. Zimin, G. Kostorz, and H. Zogg, Phys. Rev. Lett. {\bf 90}, 026104 (2003).
\bibitem{Connor05}\'E. O'Connor, A. O'Riordan, H. Doyle, S. Moynihan, A. Cuddihy, and G. Redmond, Appl. Phys. Lett. {\bf 86}, 201114 (2005).
\bibitem{Keuleyan11}S. Keuleyan, E. Lhuillier, V. Brajuskovic, and P. Guyot - Sionnest, Nature Photon. {\bf 5}, 489 (2011).
\bibitem{Rogach99}A. Rogach, S. V. Kershaw, M. Burt, M. T. Harrison, A. Kornowski, A. Eychm\"uller, and H. Weller, Adv. Mater. {\bf 11}, 552 (1999).
\bibitem{Lhuillier12}E. Lhuillier, S. Keuleyan, and P. Guyot-Sionnest, Nanotechnology {\bf 23}, 175705 (2012).
\bibitem{Keuleyan11b}S. Keuleyan, E. Lhuillier, and P. Guyot-Sionnest, J. Am. Chem. Soc. {\bf 133}, 16422 (2011).

\bibitem{Li15}J. Li, C. He, L. Meng, H. Xiao, C. Tang, X. Wei, J. Kim, N. Kioussis, G. Malcolm Stocks, and J. Zhong, Scientific Reports {\bf 5}, 14115 (2015).
\bibitem{Bernevig06}B. A. Bernevig, T. L. Hughes, and S. C. Zhang, Science {\bf 314}, 1757 (2006).
\bibitem{Konig07}M. K\"onig, S. Wiedmann, C. Br\"une, A. Roth, H. Buhmann, L. W. Molenkamp, X. L. Qi, and S. C. Zhang, Science {\bf 318}, 766 (2007).

\bibitem{Zheng15}H. Zheng, X. B. Li, N. K. Chen, S. Y. Xie, W. Q. Tian, Y. Chen, H. Xia, S. B. Zhang, and H. B. Sun, Phys. Rev. B {\bf 92}, 115307 (2015).
\bibitem{Chakraborty18}T. Chakraborty, A. Manaselyan, M. Barseghyan, and D. Laroze, Phys. Rev. B {\bf 97}, 041304(R) (2018).
\bibitem{Yeyati95}A. Levy Yeyati and M. B\"uttiker, Phys. Rev. B {\bf 52}, R14360(R) (1995).
\bibitem{Alfonso05}A. Bruno-Alfonso and  A. Latg\'e, Phys. Rev. B {\bf 71}, 125312 (2005).
\bibitem{Grbic08}B. Grbi\'c, R. Leturcq, T. Ihn, K. Ensslin, D. Reuter, and A.Wieck, Physica E {\bf 40}, 1273 (2008).
\bibitem{Konig06}M. K\"onig, A. Tschetschetkin, E. M. Hankiewicz, Jairo Sinova, V. Hock, V. Daumer, M. Sch\"afer, C. R. Becker, H. Buhmann, and L. W. Molenkamp, Phys. Rev. Lett. {\bf 96}, 076804 (2006).
\bibitem{Konig07b}M. K\"onig, H. Buhmann, C. R. Becker, and L. W. Molenkamp, Phys. Status Solidi C {\bf 4}, 3374 (2007).
\bibitem{Liang15}L. Liang and W. Xie, Physica B. {\bf 462}, 15 (2015).
\bibitem{Cantele01}G. Cantele, D. Ninno, and G. Iadonisi, Phys. Rev. B {\bf 64}, 125325 (2001).
\bibitem{Holtkemper18}M. Holtkemper, D. E. Reiter, and T. Kuhn, Phys. Rev. B {\bf 97}, 075308 (2018).
\bibitem{Zhu97}J. L. Zhu, Z. Q. Li, J. Z. Yu, K. Ohno, and Y. Kawazoe, Phys. Rev. B {\bf 55}, 15819 (1997).

\bibitem{Slater54}J. C. Slater and G. F. Koster, Phys. Rev. {\bf 94}, 1498 (1954).
\bibitem{Shen12}S. Q. Shen, {\it Topological Insulators: Dirac Equation in Condensed Matters}, (Springer, Berlin, 2012), pp. 29-31.
\bibitem{Chadi77}D. J. Chadi, Phys. Rev. B {\bf 16}, 790 (1977).
\bibitem{Allan12}G. Allan and C. Delerue, Phys. Rev. B. {\bf 86}, 165437 (2012).
\bibitem{Klein52}M. Klein, Am. J. Phys. {\bf 20}, 65 (1952).
\bibitem{Graf95}M. Graf and P. Vogl, Phys. Rev. B {\bf 51}, 4940 (1995).
\bibitem{Dumitrica98}T. Dumitric\u{a}, J. S. Graves, and R. E. Allen, Phys. Rev. B {\bf 58}, 15340 (1998).
\bibitem{Pertsova15}A. Pertsova, C. M. Canali, and A. H. MacDonald, Phys. Rev. B {\bf 91}, 075430 (2015).
\bibitem{Chen19}Q. Chen, L. L. Li, and F. M. Peeters, J. Appl. Phys. {\bf 125}, 244303 (2019).
\bibitem{Boykin01b}T. B. Boykin, R. C. Bowen and G. Klimeck, Phys. Rev. B {\bf 63}, 245314 (2001).
\bibitem{Shanavas15}K. V. Shanavas and S. Satpathy, Phys. Rev. B {\bf 91}, 235145 (2015).
\bibitem{MoldovanPB}D. Moldovan and F. M. Peeters, Pybinding v0.8.0: a python package for tight-binding calculations. (doi:10.5281/ zenodo.56818).
\bibitem{Yates07}J. R. Yates, X. Wang, D. Vanderbilt, and I. Souza, Phys. Rev. B {\bf 75}, 195121 (2007).
\bibitem{Li17}L. L. Li, D. Moldovan, W. Xu, and F. M. Peeters, Nanotechnology {\bf 28}, 085702 (2017).
\bibitem{Abdelsalam16}H. Abdelsalam, M. H. Talaat, I. Lukyanchuk, M. E. Portnoi, and V. A. Saroka, J. Appl. Phys. {\bf 120}, 014304 (2016).
\bibitem{Jakovljevic17}D. Jakovljevi\'c, M. Gruji\'c, M. Tadi\'c, and F. Peeters, J. Phys.: Condens. Matter {\bf 30}, 035301 (2017).
\bibitem{Grujic11}M. Gruji\'c, M. Zarenia, A. Chaves, M. Tadi\'c, G. A. Farias, and F. M. Peeters, Phys. Rev. B. {\bf 84}, 205441 (2011).


\end{thebibliography}
\end{document}